\documentclass{vgtc}                          % final (conference style)
\graphicspath{{figures/}{pictures/}{images/}{./}} % where to search for the images

\usepackage{times}                     % we use Times as the main font
\usepackage{tabu}                      % only used for the table example
\usepackage{booktabs}                  % only used for the table example
\usepackage{lipsum}                    % used to generate placeholder text
\usepackage{mwe}                       % used to generate placeholder figures

\usepackage{mathptmx}                  % use matching math font

\usepackage{fullpage}
\usepackage{graphicx}
\usepackage{cite}
\usepackage{url}
\usepackage{setspace}
\usepackage{fancyhdr}
\usepackage{ifthen}
\usepackage{listings}
\usepackage[section]{placeins}
\usepackage{xtab}
\usepackage{url}
\usepackage{amsmath} 
\usepackage{amssymb}
\usepackage{algorithm}
\usepackage{algpseudocode}
\usepackage{rotating} % provides sidewaystable and sidewaysfigure
\usepackage{adjustbox} % For resizing tables
\usepackage{appendix}
\usepackage{xurl}
\usepackage{multirow}
\usepackage{xcolor}

\usepackage{longtable}
\usepackage{hyperref}
\usepackage{float}
\usepackage{fancyvrb}
\usepackage{verbatim}

\lstdefinestyle{mystyle}{
    language=Python,
    backgroundcolor=\color{white},
    commentstyle=\color{purple},
    keywordstyle=\color{blue},
    numberstyle=\tiny\color{gray},
    stringstyle=\color{purple},
    basicstyle=\ttfamily\footnotesize,
    breakatwhitespace=false,
    breaklines=true,
    captionpos=b,
    keepspaces=true,
    numbers=none,
    numbersep=5pt,
    showspaces=false,
    showstringspaces=false,
    showtabs=false,
    tabsize=2
}

\onlineid{0}

\vgtccategory{Research}

\title{Can AI agents understand spoken conversations about data visualizations in online meetings?}

\author{Rizul Sharma\thanks{e-mail: sharmarz@mail.uc.edu}\\ %
    \parbox{1.4in}{\scriptsize \centering DaVInCi Laboratory \\ University of Cincinnati} %
\and Tianyu Jiang\thanks{e-mail: jiangt2@ucmail.uc.edu}\\ %
    \scriptsize University of Cincinnati %
\and Seokki Lee\thanks{e-mail: lee5sk@ucmail.uc.edu}\\ %
    \scriptsize University of Cincinnati %
\and Jillian Aurisano\thanks{e-mail: aurisajm@ucmail.uc.edu}\\ %
    \parbox{1.4in}{\scriptsize \centering DaVInCi Laboratory \\ University of Cincinnati}}

\abstract{
  In this short paper, we present work evaluating an AI agent's understanding of spoken conversations about data visualizations in an online meeting scenario. There is growing interest in the development of AI-assistants that support meetings, such as by providing assistance with tasks or summarizing a discussion. The quality of this support depends on a model that understands the conversational dialogue. To evaluate this understanding, we introduce a dual-axis testing framework for diagnosing the AI agent's comprehension of spoken conversations about data. Using this framework, we designed a series of tests to evaluate understanding of a novel corpus of 72 spoken conversational dialogues about data visualizations. We examine diverse pipelines and model architectures, LLM vs VLM, and diverse input formats for visualizations (the chart image, its underlying source code, or a hybrid of both) to see how this affects model performance on our tests. Using our evaluation methods, we found that text-only input modalities achieved the best performance (96\%) in understanding discussions of visualizations in online meetings. 

} % end of abstract


\begin{document}


\firstsection{Introduction}

\maketitle

Data-driven projects are often collaborative and interdisciplinary.  To manage these collaborations, teams periodically meet to present preliminary results, build a common understanding of the project, integrate varied perspectives and make decisions. These meetings often occur virtually, utilizing video conferencing platforms such as Zoom or Teams.  Such platforms are increasingly integrating AI-agents that utilize large-language models (LLMs) to support and summarize the discussion. For these agents to be successful, they must demonstrate a deep and accurate comprehension of what meeting participants are expressing to each other. For discussions about data, a critical challenge is ensuring that these agents understand comments about data visualizations presented in the meeting. Visualizations of data are vital tools for communicating data in meetings, providing a concrete reference point for the conversation. During a discussion, meeting participants might identify a trend or outlier in the visualization ("What is that peak in the middle?"), ask questions about data provenance ("Is this a recent sample or the entire dataset?  How was it sampled?"), or express feedback about how the data is represented visually ("The colors are hard to distinguish in that cluster.").  To understand these comments, AI-agents need to integrate content from multiple sources (transcript, visualization, data properties) and potentially multiple modalities (unstructured text, code, images, or a hybrid of all).  

There is prior work evaluating AI agents (LLMs or VLMs) in visualization literacy benchmarks \cite{han2023chartllamamultimodalllmchart,pandey2025benchmarkingvisuallanguagemodels,2025_DoLLMsHaveVisua}. In addition, researchers have evaluated language models for Chart Question Answering \cite{Dai_2024}. These assess how well these models understand questions about data visualizations. However, these evaluations do not address LLM's understanding of visualizations in relation to conversational dialogues, i.e., online meeting scenarios. There has been work to evaluate different natural language techniques for conversations about data visualizations in meeting scenarios~\cite{kumar2017towards,kumar2020augmenting,tabalba2025pragmatics}.  However, these approaches focused on classifying utterances and responding to visualization requests, instead of directly assessing language-model understanding of the discussion. 

A critical challenge is determining the most effective way to provide necessary context about a data visualization to AI agents. Relying solely on a chart image risks factual errors from imperfect visual interpretation \cite{Hiippala:2020}, while relying solely on textual context, like source code, may lead to missed visual nuances \cite{xu2025improvediterativerefinementcharttocode}. Although researchers are interested in unifying these modalities for modern AI agents, there is insufficient evidence comparing these input strategies, especially for our target context of online meetings involving data visualizations. 
% This highlights a critical grounding problem: how can we ensure the quality of an AI's visual interpretation?

%If a model relies solely on a chart image, it may make factual errors from imperfect visual interpretation \cite{Hiippala:2020}, but if it relies solely on textual content like source code, it may miss crucial visual nuances \cite{xu2025improvediterativerefinementcharttocode}.
 
% If a model relies solely on a chart image, it may make factual errors from imperfect visual interpretation, but if it relies solely on textual content like source code, it may miss crucial visual nuances

In this paper, we consider 1) how to measure an AI agent's capacity to demonstrate an understanding of spoken conversations about data visualizations and 2) what input modalities, pipelines, and model architectures best promote this understanding.  We contribute: 
\begin{itemize}
 \vspace{-2mm}
\item A corpus of 72 conversational discussions about data visualizations in an online meeting scenario.  \vspace{-2mm}
\item A benchmark of 318 questions about the discussions in the corpus. \vspace{-2mm}
\item A dual-axis evaluation framework for analyzing models' performance on the above benchmark.  \vspace{-2mm}
\item Results from a four-way comparative evaluation of input formats for visualizations (Image, Code(Text), Hybrid) and model architecture (LLM vs. VLM), using our evaluation benchmark.  \vspace{-2mm}
\end{itemize}

We found that the text-only LLM pipeline performed the best, achieving a nearly perfect accuracy of 95.9\%, followed by the text-only VLM pipeline, achieving 72.4\% accuracy. The hybrid pipeline performed the worst with an accuracy score of 68\%, while the image-only pipeline scored 70\%. We discuss the implications of these findings for the development of AI-support tools for online meetings about data.  

% \section{Related Work}

\section{Online Meeting Data Corpus}

% To evaluate AI agents' understanding of spoken conversations about visualizations in meetings, we conducted a study to build a corpus. 
First, we needed to build a corpus of conversations about data visualizations in an online meeting scenario.  This corpus is used to test AI-agent understanding.  

% \subsection{Recruitment and Protocol}

\textbf{Recruitment and Protocol:}  We recruited participants who had some experience working with data. These participants were recruited from the University mailing lists.  This study took place over Zoom, with two participants and one researcher. The researcher showed the participants a set of slides depicting 8 data visualizations. These visualizations included bar charts, line charts, histograms, box-plots, and scatter plots. These charts visualized a movie dataset \cite{rounakbanik_movies_dataset}, which is originally sampled from the MovieLens dataset~\cite{Harper:2015} utilizing the TMDB Open API. We selected this dataset because we believed it to be familiar to a general audience. During the session, the researcher introduced each visualization. Meeting participants were encouraged to discuss the data and visualization freely. Each session was recorded and transcribed using Zoom. Following the sessions, we divided each transcript into 8 sub-parts for each of the visualizations discussed.

% \subsection{Corpus description}

\textbf{Corpus description: } We recruited 18 participants (16M, 2F) to 9 paired mock-meeting sessions.  Participant ages ranged from 21 to 35 years old, from backgrounds such as computer science, business administration, biostatistics, and finance. Sessions ranged from 17 to 55 minutes, with transcripts of 2228 to 7343 words. This produced 72 (8*9) transcript-chart pairs.  We conducted two pilot sessions with 4 participants to refine our study protocol.

\section{Benchmark questions}

Our goal is to develop a method that assesses how well AI-pipelines showcase an understanding of spoken conversational discussions about data visualizations. Two researchers iteratively designed a set of 318 evaluation questions in total for the 72 transcript-chart pairs. These questions were designed to cover significant discussion points in each conversation. They also referenced content present in the transcript and visualization. These questions were posed in a multiple-choice format to allow for automated scoring of pipeline performance, mirroring prior language-model evaluation approaches~\cite{han2023chartllamamultimodalllmchart,pandey2025benchmarkingvisuallanguagemodels}.  

%Two researchers iteratively designed a set of 318 evaluation questions for each of the 72 transcript-chart pairs.
% The benchmark questions are included in the supplemental materials and is published online as described in Section 9.

\section{Dual Axis Evaluation Framework}

To provide a more granular score of model performance, we labeled our questions based on 1) level of complexity and 2) topic. These two labels formed our dual-axis framework for diagnosing LLM comprehension. The distribution of the 318 benchmark questions on the two axes is available in the supplementary materials (Section 9).

\subsection{Level of complexity} 

Questions were sorted into 3 categories of increasing complexity. These labels are inspired by Bloom's Taxonomy, which is a hierarchical framework for categorizing learning objectives from simpler tasks,
such as basic recall of factual information, up to more complex activities like synthesis and evaluation of information~\cite{bloom1956taxonomy}. This taxonomy is often used by instructors to design assessments that measure student learning at different levels of complexity. We selected Bloom's taxonomy as a reference for our assessment because we wished to measure our pipeline's understanding in the same way you might assess the understanding of a student.  Our three complexity labels are as follows: 
\textbf{Level 1: Factual Recall}: Does the model showcase a capacity to report the basic facts of what was said and what happened? Example questions in this category may involve asking the model to recall the insights or points directly expressed by participants in the dialogue. 
\textbf{Level 2: Data Interpretation}: Can the model identify the trends, patterns, and outliers in the data alluded to or implied by participants in the discussion? Example questions in this category require connecting opinions or observations to the underlying pattern in the data. 
\textbf{Level 2: Participant State Analysis}: Can the model go beyond the recall of statements to gauge opinions and understand reactions? Example questions in this category require inferring opinions or beliefs from participant statements. 
\textbf{Level 3: Causal and Process Analysis}: Can the model explain the underlying reasons for participant statements about their opinions? Example questions ask the model to identify hypotheses and theories behind participant statements.

% \textbf{Level 2: Interpretive Comprehension (Data Interpretation \& Participant State Analysis)}
% Can the model go beyond the facts to understand what those facts mean? This tests its ability to interpret data, gauge opinions, and understand reactions.
% \textit{Example question for Data Interpretation:} What was Participant 6's primary conclusion based on the general trend shown in the scatter plot?
% \textit{Example question for Participant State Analysis:} What was the participants' shared opinion about using two-letter abbreviations?

Although Bloom's taxonomy features six levels \cite{bloom1956taxonomy}, we selected three based on applicability to our tasks. Our 'Level 1: Factual recall' label directly maps to the first level, "remembering". Our 'Level 2: Data Interpretation, Participant State Analysis' labels encompass the cognitive skills of "understanding" and "applying" concepts to the data. Finally, our 'Level 3: Causal and Process Analysis' label corresponds to "analyzing", as it requires the model to deconstruct participant arguments to understand causal theories. The higher levels of the taxonomy, "evaluating" and "creating", are more aligned with generative tasks, which we will pursue in future work.
% whereas our evaluation questions are designed to test the agent's understanding of the human-led analysis that occurred in the conversation.

\subsection{Topics}

To form the second dimension of our dual-axis framework, we designed a set of topic tags to pinpoint the visualization elements or discussion points referenced in the question. The tags are designed for the questions about participant statements related to 
\textbf{Chart Layout:} Overall visualization, such as bars being too big or too small;
\textbf{Axis \& Labels:} X-axis, Y-axis, or their labels;
\textbf{Visual Encoding:} How data is encoded, such as color scales;
\textbf{Data Provenance:} The origin, sample size, time period, or accuracy of the data;
\textbf{Data Pattern:} Identified trends, outliers, and patterns in the data; 
\textbf{Data Point:} Focusing on a single specific data point under discussion;
\textbf{Participant Hypothesis:} A specific explanation or reason, such as a participant explaining a trend or pattern in the data.

\section{Comparative Pipeline Design}

We used this evaluation method to compare four different pipelines, which utilized different models and visualization input formats: (1) a Vision Language Model (VLM) which is given a visualization image as input, (2) an Large Language Model (LLM) which is given the code for the visualization, (3) a VLM given the code for the visualization (tested to isolate architecture from input format), and (4) a VLM analyzing a Hybrid of both image and text (code) for the visualization input. We selected llava-1.5-7b-hf \cite{liu2024improvedbaselinesvisualinstruction} for the VLM and Mixtral-8x7B-Instruct-v0.1 \cite{jiang2024mixtralexperts} for the LLM, as both fit our testing requirements for the two different model architectures and allowed us to test our core questions related to image, text, and hybrid inputs without the confounding factor of extreme latency or memory-related failures.

\begin{figure}
    \centering
    \includegraphics[width=\linewidth]{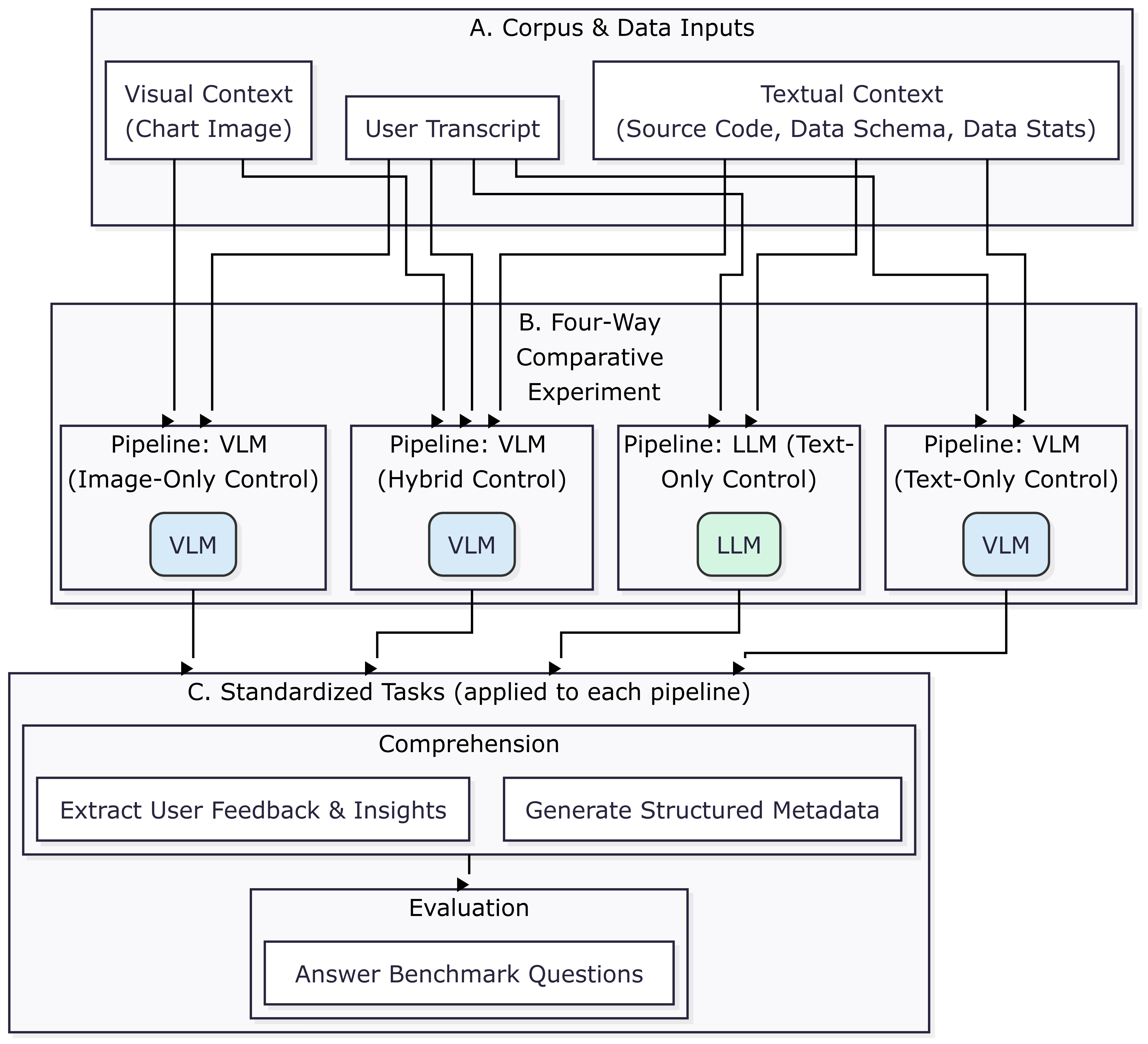}
    \caption{High-level flow for pipeline operations: (A) Inputs (B) Four Pipeline Modes (C) Standardized Comprehension Tasks }
    \label{fig:enter-label}
\end{figure}

For each pipeline, the AI agent performs a series of initial steps to extract information from its input. This includes extracting user insights/feedback from the transcripts, generating structured metadata from the visualization images (for the VLM-Image and VLM-Hybrid pipelines). We then pose the questions to test each pipeline's comprehension of the user discussions, as shown in Figure~\ref{fig:enter-label}.

% \subsection{Analysis methods}

\textbf{Analysis methods:} Quantitative analysis is performed by calculating the accuracy of each pipeline's answers to the benchmark questions, broken down by complexity and topic, as discussed previously. This approach enables us to analyze specific strengths and weaknesses. For instance, we can isolate and examine all Level 2 (Interpretive) errors that fall under the 'Data Provenance' topic tag to understand why a model struggles with that specific task.

\section{Findings}

% \subsection{Pipeline Performance Overview}

The Text-only LLM pipeline performed the best, achieving a nearly perfect accuracy of 95.9\%, followed by the Text-only VLM pipeline, achieving 72.4\% accuracy. This shows that even with the same input modality, the architecture of the pipeline, in this case, an LLM versus VLM, can significantly affect its performance. The Hybrid pipeline performed the worst with an accuracy score of 68\%, while the Image-only pipeline scored 70\% on our evaluation. This was an interesting observation since the Hybrid pipeline had more contextual knowledge input compared to the other pipelines.

\begin{figure}
    \centering
    \includegraphics[width=\linewidth]{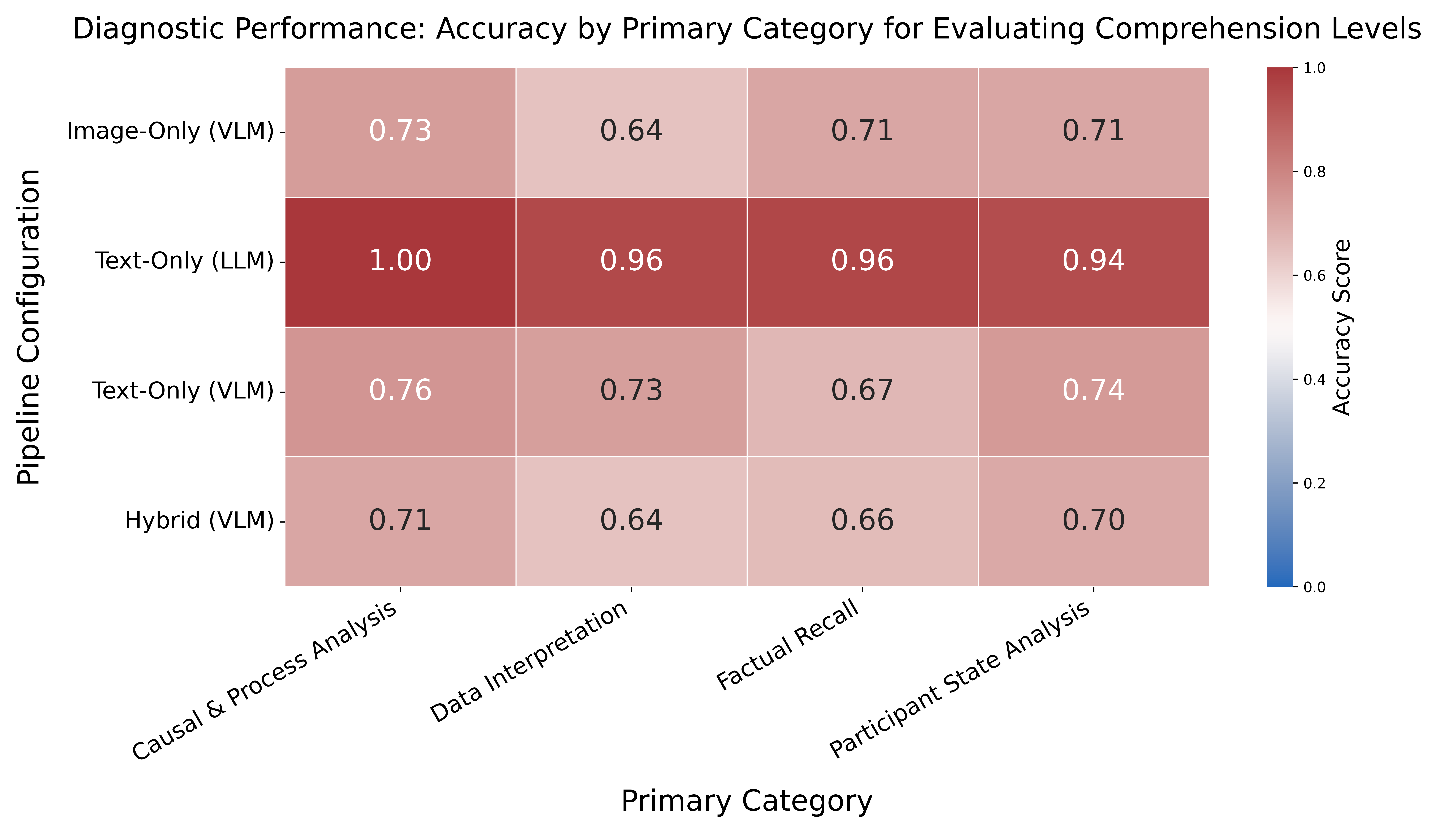}
   \includegraphics[width=\linewidth]{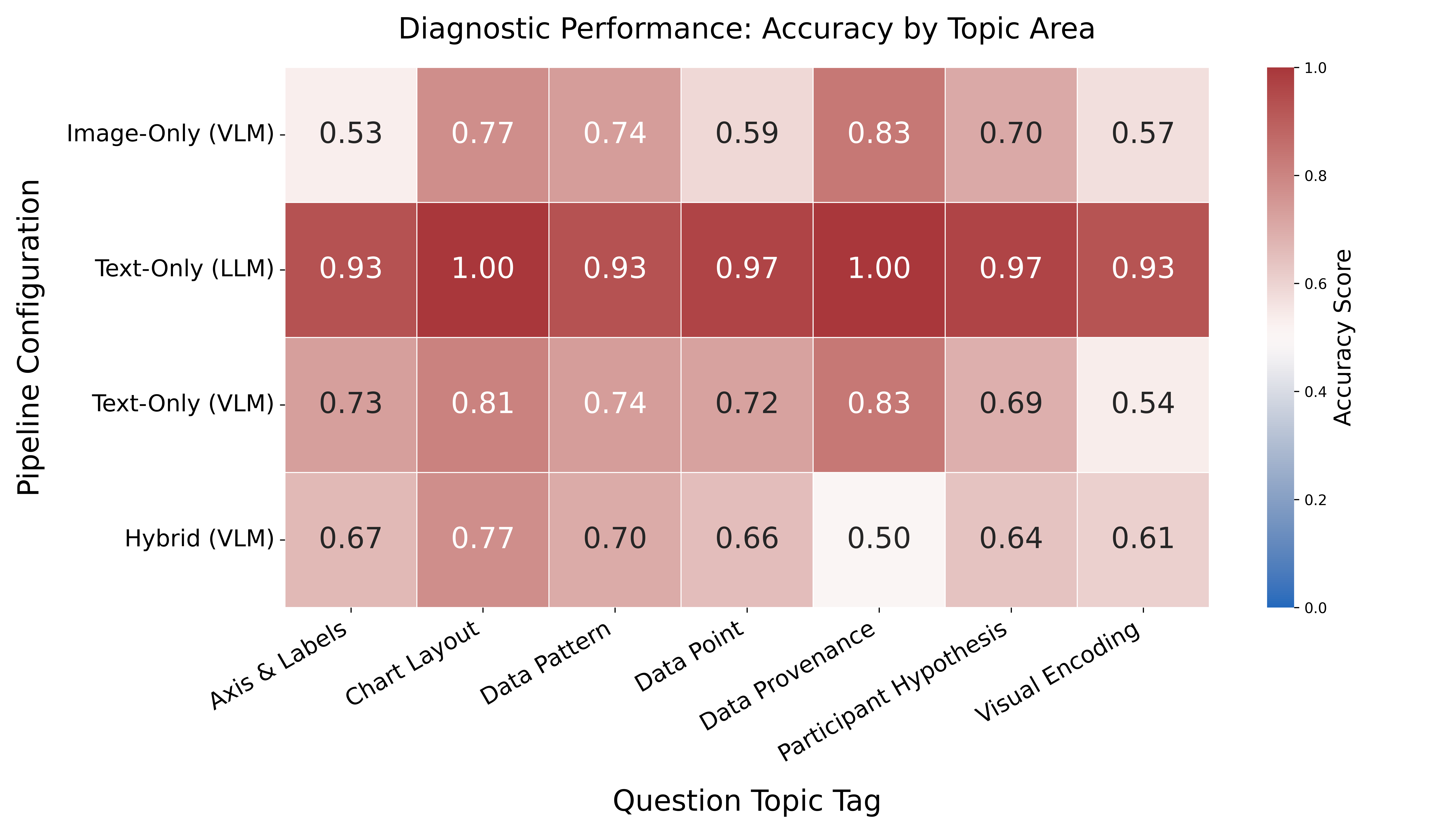}    \caption{(top) Heatmap - Performance on Complexity Levels. (bottom) Heatmap - Performance on Topic Tags}
    \label{fig:Diag_prim}
\end{figure}

%\begin{figure}
%    \centering
%    \includegraphics[width=\linewidth]{figures/diagnostic_performance_heatmap.png}
%    \caption{Heatmap - Performance on Areas of Weakness}
%    \label{fig:diag_topic}
%\end{figure}

% \subsection{Text-Only (LLM): The Gold Standard}

\textbf{Text-Only (LLM): The Gold Standard} This pipeline is the top performer with a score greater than or equal to 90\% in almost every category for every comprehension level (Figure 2 t,b). The only areas where it scored less than 90\% were Axis \& Labels for Data Interpretation (67\%), Data Pattern for Factual Recall (81\%), and Visual Encoding for Participant State Analysis (80\%). These results suggest that with structured text input (source code, schema), the LLM can robustly handle all levels of comprehension, from factual recall to analysis.

% % % % \begin{figure}[htbp]
% % % %     \centering
% % % %     \fbox{\includegraphics[width=\linewidth]{figures/category_vs_topic_analysis_text_only.png}}
% % % %     \caption{Heatmap - Text-only (LLM) Accuracy Analysis Evaluation Matrix}
% % % %     \label{fig:text_only}
% % % % \end{figure}

% \subsection{Text-Only (VLM): The Architectural Contender}
\textbf{Text-Only (VLM): The Architectural Contender}  As discussed earlier, this pipeline was designed to isolate the model architecture as a variable.  Although it performs reasonably well, its accuracy of 72.4\% is significantly lower and more inconsistent than the 95.9\% accuracy of the Text-Only (LLM) pipeline, despite receiving the exact same text-only input. For example, it struggles with Causal \& Process Analysis on Visual Encoding (20\%) and general Factual Recall. This suggests that the architecture of the underlying model has a tangible impact on reasoning ability. The LLM is demonstrably better at text-based reasoning tasks than the VLM, even when the VLM is not "distracted" by an image in this case.

% % % \begin{figure}[htbp]
% % %     \centering
% % %     \fbox{\includegraphics[width=\linewidth]{figures/category_vs_topic_analysis_llava_text_only.png}}
% % %     \caption{Heatmap - Text-only (VLM) Accuracy Analysis Evaluation Matrix}
% % %     \label{fig:llava_text_only}
% % % \end{figure}

% \subsection{Hybrid (VLM): The Confused Performer}
\textbf{Hybrid (VLM): The Confused Performer} This pipeline's performance indicates challenges in fusing inputs (text \& image). Although it performs well on some tasks (e.g., Causal \& Process Analysis on Axis \& Labels and on Chart Layout), it fails on others. For example, it achieved zero accuracy in Data Interpretation on Data Provenance questions, on which the Text-Only VLM scored perfectly, which implies that adding the chart image caused a performance collapse. The model may have exhibited a knowledge conflict, ignoring the correct textual information in favor of incorrect visual information.

% % \begin{figure}[htbp]
% %     \centering
% %     \fbox{\includegraphics[width=\linewidth]{figures/category_vs_topic_analysis_hybrid.png}}
% %     \caption{Heatmap - Hybrid (VLM) Accuracy Analysis Evaluation Matrix}
% %     \label{fig:hybrid}
% % \end{figure}

% \subsection{Image-Only (VLM): The Brittle Visionary}
\textbf{Image-Only (VLM): The Brittle Visionary}  This pipeline demonstrates the unreliability of relying solely on visual interpretation for complex analysis. The scores vary significantly. It performs perfectly on some high-level tasks (for example, Causal \& Process Analysis on Chart Layout - 100\%), but fails on others (for example, Causal \& Process Analysis on Axis \& Labels - 0\%). It struggles with tasks requiring precise data extraction, such as Data Interpretation on Data Point (40\%) and Axis \& Labels (33\%). This suggests that while it might grasp the general layout, it cannot be trusted for detailed, factual analysis and fine-grained visual perception.

% \begin{figure}[htbp]
%     \centering
%     \fbox{\includegraphics[width=\linewidth]{figures/category_vs_topic_analysis_image_only.png}}
%     \caption{Heatmap - Image-only (VLM) Accuracy Analysis Evaluation Matrix}
%     \label{fig:image_only}
% \end{figure}

%\input{tables/ramones_stats}

% It will be interesting to test this on an expanded corpus and benchmark.
 
\section{Discussion}

% This paper aimed to determine the most effective way to input modality and determine the architecture for an AI agent to understand spoken conversations about data visualizations. Our findings from a four-way comparative experiment indicate that providing structured textual context, such as the visualization's source code and data schema, is the most reliable method for grounding an AI and achieving relatively high comprehension. Pipelines relying on visual interpretation were fragile and prone to errors \cite{kolner2025mindgapglimpsebasedactive}, while the hybrid pipeline suffered from knowledge conflicts, often trusting its flawed visual analysis over the provided ground-truth text \cite{jia2025benchmarkingmultimodalknowledgeconflict}.
Our findings suggest that AI agents perform well in our benchmark when given structured textual inputs, such as visualization code and data schemas. Given a 96\% performance of the text-only LLM pipeline on our benchmark, there are strong reasons to believe that AI-agents with an LLM can be explored as support tools for meetings about data.  However, these findings do not imply limiting inputs to text, but encourage researchers to explore better strategies for combining modalities.

\textbf{Implications:}  These findings encourage us to explore AI's potential collaboration in meetings. Future studies should assess how the presence and interventions of an AI agent impact the nature of human-to-human conversation; the aim should be to make AI a helpful collaborator instead of an unwelcome intruder. Our findings also highlight a challenge for the design of AI-agents for meetings. Typical tools for online meetings, such as Slideware like PowerPoint, and video conferencing systems like Zoom \& Teams, primarily have access to visualizations in the form of images, either embedded in the slide or within the video feed. However, we found that image-only information about visualizations may be insufficient for achieving model understanding of the discussion. Meeting tools may need to embed visualization code and other metadata in text form to enable effective AI support.

\textbf{Limitations:} Our current corpus is focused on participants with a range of data visualization experience examining charts from a tabular dataset. It is possible that we would find different performance results given a corpus of conversations between experts in various domains with spatial, temporal, or graph-structured data in different visual templates. In addition, our evaluation approach required human-generated benchmark questions, which may be difficult to extend and scale for more domains.  Finally, given the strong performance of the LLM on our benchmark, future benchmarks may need to be more challenging and based on more complex and domain-specific corpora, so that we can distinguish the performance improvements of future model pipelines.

%Solve real-world challenges in online meetings, for example, research in topic drift detection
%grounding

\section{Conclusion}

We contribute a novel corpus of 72 conversational discussions about visualizations in an online meeting setting. We developed a benchmark of 318 questions to evaluate the AI-pipelines' understanding of these discussions. We created a dual-axis framework \& labeled these questions based on complexity and topic. Finally, we compared four pipelines with different model architectures (VLM vs LLM) and input modalities (text, image, and hybrid), using our benchmark and corpus.  We found that text-only LLM achieved 96\% accuracy on our benchmark. Future environments for online meetings may need to consider ways to integrate text-based information about visualizations to design AI-driven support tools.

\section{Supplemental Materials}

To promote future benchmarking and reproducibility, materials from this paper are published at \href{https://github.com/anij715/AI-multimodal-meeting-assistant/tree/main}{GitHub} \cite{Sharma_AI-multimodal-meeting-assistant_A_Multimodal_2025}, including: (1) code for the pipelines, (2) study visualizations, (3) the corpus of discussions, and (4) the benchmark questions and labels.

% All listed materials are the intellectual property of the authors.

%\section*{Figure Credits}
%\label{sec:figure_credits}

%Refer to the instructions for this section (\cref{sec:figure_credits_inst}).
%Here are the actual figure credits for this template:

%\Cref{fig:teaser} image credit: Scott Miller / Special to the Vancouver Sun, January 22, 2009, page A6.

%\Cref{fig:vis_papers} is a partial recreation of Fig.\ 1 from \cite{Isenberg:2017:VMC}, which is in the public domain.

% The authors wish to thank A, B, and C. This work was supported in part by
% a grant from XYZ.}

%\bibliographystyle{abbrv}
\bibliographystyle{abbrv-doi}

\bibliography{template}
\end{document}